\documentstyle[12pt]{article}
\begin{document}

It is obvious that new physics is conducted now and willcontinue in rare
processes in a foreseen future.Prior to their direct detection on colliders,
supersymmetry effects may manifest themselves implicitly in rare processes.
Supersymmetry gives elegant solutions to the problems of the standard model and
predicts (minimal supersymmetry) the masses of supersymmetry particles in the
range $100 Gev - O(1) TeV$ (with the exception of the lightest supersymmetry
particle). In the present paper we consider the $B^{0}$-meson's rare
double-photon radiative decays $B^{0} \rightarrow \gamma\gamma$ in the
framework of the minimal supersymmetric standard model(MSSM).
    The process   $B_{s} \rightarrow \gamma\gamma$ is similar to the decay
$K^{0} \rightarrow \gamma\gamma$.
However in the former process long-distance effects does not play an important
role and short-distance contributions are the leading one. It should be
mentioned
that the two photons in the final state can be both in CP-odd and CP-even
configuration. This possibility could become as a nontraditional source for
the search for CP-violating effects in B-physics.

The investigation of the $B$-meson's rare decays is of great interest in
order to test the standard model and beyond standard model physics.
It seems that experimental investigation of beauty physics will be one
of the major topics at the world facilities.
The L3 collaboration(CERN) established that[1]
$$
Br(B_{s} \rightarrow\gamma\gamma) < 14,8 \cdot 10^{-5}
$$
$$
Br(B_{d} \rightarrow\gamma\gamma) < 3,95 \cdot 10^{-5}
\eqno{(1)}
$$
The experimental interest stimulates theoretical investigations [2-5].
All group of the authors have obtained (in the frame of the standard model)
the same values for the branching ratios in the leading order $1/M_{W}^{2}$ :
$$
Br(B_{s} \rightarrow \gamma\gamma)=(3.0 \pm 1.0) \cdot 10^{-7}
$$
$$
Br(B_{d} \rightarrow \gamma\gamma)=(1.2 \pm 1.1) \cdot 10^{-8}
	      \eqno{(2)}
$$

In the framework of supersymmetric extension of the
standard model one has following set of diagrams which contribute into decays
$B \rightarrow \gamma\gamma$
(the diagrams are classified by particles
which appear in the loops): a) charged gauge fermions(chargino) and up scalar
quarks, b) charged Higgs particles and up quarks, c) neutral gauge
fermions(neutralino) and down scalar quarks, d) gluino and down scalar quarks.
The Lagranjian of interaction down quarks witth supersymmetric
particles has the following form[6]:
$$
  L_{d{\tilde{\chi}}^{*}{\tilde{u}}}=-\frac{ig}{2}[V_{j1}(1 + \gamma_{5}) -
\frac{m_{d}V_{j2}^{*}}{\sqrt{2} M_{W}\cos\beta}(1 - \gamma_{5})]C
\Gamma^{L}_{ab}{\bar{d}}_{a}{\tilde{\chi}}^{*}{\tilde{u}}_{Lb} +
$$
$$
\frac{igm_{u}V_{j2}}{2\sqrt{2}M_{W}\sin\beta}(1 - \gamma_{5})C\Gamma^{R}_{ab}
{\bar{d}}_{a}{\tilde{\chi}}^{*}{\tilde{u}}_{Rb},
$$
$$
L_{dHu} = \frac{ig}{2\sqrt{2} M_{W}}[m_{d}\tan\beta (1 - \gamma_{5}) +
 m_{u}\cot\beta (1 + \gamma_{5})]O_{ab}{\bar d}_{a}Hu_{b}
$$
$$
  L_{d{\tilde{\chi}}^{0}{\tilde{u}}}=-\frac{i}{\sqrt{2}}
\bigl\{ \frac{gm_{d}}{2M_{W}\cos\beta}N_{j3}^{*}(1 - \gamma_{5})  +
[eQ{_d}N_{j1} -
$$
$$
\frac{g}{\cos\theta_{W}}(\frac12 + Q_{d}\sin^{2}\theta_{W})N_{j2}]
(1 + \gamma_{5})\bigr\}
F^{L}_{ab}{\bar{d}}_{a}{\tilde{\chi}}^{0}{\tilde{u}}_{Lb} -
$$
$$
\frac{i}{\sqrt{2}}
\bigl\{ \frac{gm_{d}}{2M_{W}\cos\beta}N_{j3}(1 + \gamma_{5})  -
[eQ{_d}N_{j1}^{*} -
N^{*}_{j2}\frac{gQ_{d}\sin^{2}\theta_{W}}{\cos\theta_{W}}]
(1 - \gamma_{5})\bigr\}
F^{R}_{ab}{\bar{d}}_{a}{\tilde{\chi}}^{0}{\tilde{u}}_{Rb}
$$
$$
L_{d{\tilde{g}}{\tilde{d}}}
 = -\sqrt{2}g_{s}\frac{\lambda^{a}_{ij}}{2}F^{L}_{ij}
\frac{1 + \gamma_{5}}{2}{\bar{d}}_{i}{\tilde{g}}_{a}{\tilde{d}}_{Lj} +
\sqrt{2}g_{s}\frac{\lambda^{a}_{ij}}{2}F^{R}_{ij}
\frac{1 - \gamma_{5}}{2}{\bar{d}}_{i}{\tilde{g}}{\tilde{d}}_{Rj} ,\eqno{(3)}
$$
where $g=e\sin\theta_{W}$,
$V$ and $U$   are charged gauge fermions' mixing matrices,
$C$ is charge conjugtion matrix,
$\Gamma^{L}, \Gamma^{R}$ are up skalar quarks mixing matrices,
$tg\beta =v_{2}/v_{1}$, $v_{1,2}$ are are scalar Higgs fields' vacuum
expectation[6], $O$ is Kabibo-Kobayashi-Maskava matrix,
$N$ is neutral gauge fermins mixing matrix [6], $Q_{d}=-1/3$, $F^{L},
F^{R}$ are down scalar quarks mixing matrices, $g_{s}$ is strong interaction
coupling constant, $\lambda^{a}$ are Gell-Mann matrices.

   One can write down the amplitude for the decays
$B \rightarrow \gamma\gamma$ in the following form, which is correct
after gauge fixing for final photons:
$$
T(B \rightarrow \gamma\gamma)=\epsilon^{\mu}_{1}(k_{1})
\epsilon^{\nu}_{2}(k_{2})[Ag_{\mu\nu} + iB\epsilon_{\mu\nu\alpha\beta}
k_{1}^{\alpha}k_{2}^{\beta}]                           \eqno{(4)}
$$
where $\epsilon_{1}^{\mu}(k_{1})$ and $\epsilon_{2}^{\nu}(k_{2})$ are
the polarization vectors of final photons with momenta $k_{1}$ and $k_{2}$
respectively.

Let us mention that our calculations are performed in the frame of
Feynman-t'Hooft gauge and we use dimension regularization technique for
divergent Feynmann diagrams. Only one particle reducible diagrams contain
divergent parts. The divergent parts mutually cancel in the sum of amplitudes
and due to the GIM mechanism[7].

Ccontribution of the diagrams with charged gauge fermions into amplitudes A and
B have the following forms:
$$
 A=i\frac{\sqrt{2}m_{b}}{16\pi^{2}m_{s}}G_{F}f_{B}(eQ_{D})^{2}M^{2}_{B}
\sum\lambda_{l}\{\frac{\sqrt{2}M_{W}}{M({\tilde{\chi}}^{*}_{j})\cos\beta}
(V_{j1}U_{j2}-\frac{m_{s}}{m_{b}}V^{*}_{j1}U^{*}_{j2})f_{1}(x_{l}) +
$$
$$
(\mid V_{j1}\mid^{2} - \frac{m_{s}m_{b}\mid U_{j2}\mid^{2}}
{2M_{W}^{2}\cos^{2}\beta})
\frac{M_{W}^{2}}{M^{2}({\tilde{\chi}}^{*}_{j})}f_{2}(x_{l})\},
$$
$$
 B=i\frac{\sqrt{2}m_{b}}{8\pi^{2}m_{s}}G_{F}f_{B}(eQ_{D})^{2}
\sum\lambda_{l}\{\frac{\sqrt{2}M_{W}}{M({\tilde{\chi}}^{*}_{j})\cos\beta}
(V_{j1}U_{j2}+\frac{m_{s}}{m_{b}}V^{*}_{j1}U^{*}_{j2})f_{1}(x_{l}) +
$$
$$
(\mid V_{j1}\mid^{2} + \frac{m_{s}m_{b}\mid U_{j2}\mid^{2}}
{2M_{W}^{2}\cos^{2}\beta})
\frac{M_{W}^{2}}{M^{2}({\tilde{\chi}}^{*}_{j})}f_{2}(x_{l})\}, \eqno{(5)}
$$
where
$$
 x_{l}=\frac{m^{2}({\tilde{u}}_{Ll})}{M^{2}({\tilde{\chi}}^{*}_{j})},~~~~~~
f_{1}(x)=\frac{5 - 12x + 7x^{2} + (4x - 6x^{2})\ell nx}{2(1 - x)^{3}},
$$
$$
f_{2}(x)=\frac{29 - 96x + 111x^{2} - 44x^{3} +
6x(4 - 9x + 6x^{2})\ell nx}{6(1 - x)^{4}}. \eqno{(6)}
$$

The higgs particles contribution into A and B
have the following forms(contributions of other supersymmetric particles are
significant small):
$$
 A=i\frac{\sqrt{2}m_{b}}{16\pi^{2}m_{l}}G_{F}f_{B}(eQ_{D})^{2}M^{2}_{B}
\sum O_{bi}O_{si}^{*}
\{f_{1}(x_{i}) +\cot^{2}\beta f_{2}(x_{i}) -
\tan^{2}\beta \frac{m_{l}m_{b}}{M^{2}(H)}\frac{1}{x_{i}}f_{2}(x_{i})\},
$$
$$
 B=i\frac{\sqrt{2}m_{b}}{8\pi^{2}m_{l}}G_{F}f_{B}(eQ_{D})^{2}
\sum O_{bi}O_{si}^{*}
\{f_{1}(x_{i}) + \cot^{2}\beta f_{2}(x_{i}) +
\tan^{2}\beta \frac{m_{l}m_{b}}{M^{2}(H)}\frac{1}{x_{i}}f_{2}(x_{i})\},
\eqno{(7)}
$$
where $f_{B}$ is the $B$-meson decay constant, $O_{ij}$ being elements of
Cabbibo-Kobayashi-Maskawa matrix, $l=s,d$, $Q_{d}=-1/3$ is charge of down
quarks, $M_{B_{s}} \approx m_{b}+m_{s}$,
$\tan\beta =v_{2}/v_{1}$, ($v_{1,2}$ being
vacuum expectation of the Higgs bosons)[6] and we have introduced following
notation:
$$
x_{i}=\frac{m^{2}(u_{i})}{M^{2}(H)},~~~~
f_{1}(x)=\frac{-3x + 8x^{2} - 5x^{3} + (6x^{2} - 4x)\ell nx}{2(1 - x)^{3}},
$$
$$
f_{2}(x)=\frac{31x - 84x^{2} + 69x^{3} - 16x^{4} +
6x(x^{2} - 6x + 4)\ell nx}{12(1 - x)^{4}}, \eqno{(8)}
$$
One can obtain contributions other supersymmetric particles
from formulae (5) by following way:\\
Charged gauge fermions and right handed up scalar quarks:
$$
V_{j1} \rightarrow 0,~~~~~~\frac{\mid U_{j2} \mid^{2}}{\cos^{2}\beta}
\rightarrow \frac{m^{2}({\tilde{u}}_{Rl})}{m_{b}m_{s}}
\frac{\mid V_{j2}\mid^{2}}{\sin^{2}\beta},
$$
$$
m(\tilde{u_{Ll}}) \rightarrow m({\tilde{u}}_{Rl}),~~~~~~\Gamma^{L}_{li}
\rightarrow \Gamma^{R}_{li},  \eqno{(9)}
$$
$$
 M({\tilde{u}}_{Ll}) \rightarrow M(H),~~~~~~
 M({\tilde{\chi}}^{*}_{j}) \rightarrow m(u_{l}), \eqno{(10)}
$$
Neutral gauge fermions and left handed down scalar quarks:
$$
V_{j1} \rightarrow{2} [\sin\theta_{W} N_{j1} -\frac{1}{\cos\theta_{W}}
(\frac{1}{2} + Q_{d}\sin^{2}\theta_{W})N_{j2}]\equiv t_{Lj} ,
$$
$$
U_{j2} \rightarrow -N_{j3},~~~~~~Q_{u}\rightarrow 0,~~~~~~
 M({\tilde{\chi}}^{*}_{j}) \rightarrow M({\tilde{\chi}}^{0}_{j}),
$$
$$
m({\tilde{u}}_{Ll}) \rightarrow m({\tilde{d}}_{Ll}),~~~~~~
\Gamma^{L}_{li} \rightarrow F^{L}_{li} , \eqno{(11)}
$$
One can obtain expressions for the diagrams with neutral gauge fermions and
righ handed down scalar quarks from the contribution of neutral gauge fermions
and left handed down scalar quarks by following way:
 \vspace{1cm}
$$
t_{Lj} \rightarrow \frac{m_{s}}{\sqrt{2}M_{W}\cos\beta}N_{j3},~~~~~
t^{*}_{Lj} \rightarrow \frac{m_{b}}{\sqrt{2}M_{W}\cos\beta}N^{*}_{j3},
$$
$$
 \frac{m_{s}}{\sqrt{2}M_{W}\cos\beta}N^{*}_{j3} \rightarrow -\sqrt{2}
(\sin\theta_{W}Q_{d}N^{*}_{j1} - Q_{d}\frac{\sin^{2}\theta_{W}}
{\cos\theta_{W}}N^{*}_{j2}) \equiv t_{Rj}^{*},
$$
$$
 \frac{m_{b}}{\sqrt{2}M_{W}\cos\beta}N_{j3} \rightarrow t_{Rj},~~~~~~
m({\tilde{d}}_{Ll}) \rightarrow m({\tilde{d}}_{Rl}),~~~~~~
\Gamma^{L}_{li} \rightarrow F^{R}_{li} , \eqno{(12)}
$$
One can obtain expressions for the diagrams with gluino
from the contributions of the neutral gauge fermions by following way:
$$
 N_{j3} \rightarrow 0,~~~~g \rightarrow \frac{4}{3} g_{s},~~~~
t_{Lj} \rightarrow \sqrt{2},~~~~{\tilde{\chi}}^{0}_{j} \rightarrow \tilde{g}
$$
$$
 N_{j3} \rightarrow 0,~~~~g \rightarrow \frac{4}{3} g_{s},~~~~
t_{Rj} \rightarrow -\sqrt{2},~~~~{\tilde{\chi}}^{0}_{j} \rightarrow \tilde{g}.
\eqno{(13)}
$$

Using equation (4) we find that the branching ratio of the decays
$B \rightarrow \gamma\gamma$ can be represented as
$$
Br(B\rightarrow \gamma\gamma)=\frac{1}{32 \pi M_{B}\Gamma_{tot}}
[4\mid A \mid^{2} + \frac{1}{2}M_{B}^{4}\mid B \mid^{2}]. \eqno{(14)}
$$
With the aid of equation (4) we can find the relative fraction of CP-odd state
of two final photons is given by
$$
  \delta=\frac{4A^{2}}{4A^{2}+(1/2)M_{B}^{4}B^{2}}   \eqno{(15)}
$$

Using formulae (4)-(15) one can estimate the branching ratio and parameter
$\delta$
of
the decay $B_{s} \rightarrow \gamma\gamma$,
We have used the following set of parameters:
$\Gamma_{tot}(B_{s})=5\cdot 10^{-4} eV$, $m_{t}=174 GeV$, $m_{b}=4,8 GeV$,
$m_{s}=0,5 GeV$,  $f_{B_{s}}=200 MeB$ [8].
For small value of the Higgs particles masses
($M(H)\approx{100GeV}$) and large $\tan\beta$ the supersymmetric contribution
into branching ratio of the deacay $B_{s}\rightarrow\gamma\gamma$ are
significant large than standard model estimate
$(Br_{susy}(B_{s}\rightarrow\gamma\gamma)\sim 10^{-6}>
Br_{SM}(B_{s}\rightarrow\gamma\gamma)\sim 10^{-7})$. In the wide range of
supersymmetric parameters $(1<tan\beta<50,~~~100 GeV < M(H) < 350 GeV)$
the supersymmetric contribution on the same level or more as standard model
estimate ($Br_{SM}(B_{s}\rightarrow\gamma\gamma)\sim 10^{-7})$.

The authors express their deep gratitude to N.~Amaglobeli, T.~Kopaleishvili,
A.~Tavkhelidze, Z.~Garuchava, T.~ Sakhelashili and M.~Xaindrava for support
and interesting discussions .
\begin{center}
References
\end{center}
1. L3 Collaboration preprint CERN-PPE-95-136\\
   L3 Collaboration, contributed paper to theEPS Conference, (EPS-0093-2)
Brussels,1995\\
2. G.-L. Lin, J. Liu, Y.-P. Yao // Phys. Rev. 1990. V. D42. P. 2314.\\
3. H. Simma, D. Wyler // Nucl. Phys. 1990. V. B344. P. 283.\\
4. S. Herrlich, J. Kalinowski // Nucl. Phys. 1993. V. B381. P. 1176.\\
5. G.G. Devidze, G.R. Jibuti, A.G. Liparteliani // Nucl. Phys. 1996. V.
B468.  P.241.\\
6. H.E. Haber, G.L. Kane // Phys.  Rep.  1985.  V.  117.  P.  75.\\
6. Glashow S. L., Illiopoulos J., Maiani L. // Phys. Rev. 1970. V. D10.
P. 897.\\
8. Review of Particle properties//Phys. Rev. 1996. V. D54.
\end{document}